\documentclass{PoS}
\newcommand{\be}{\begin{equation}}  
\newcommand{\ee}{\end{equation}}  
\newcommand{\beq}{\begin{eqnarray}}  
\newcommand{\eeq}{\end{eqnarray}}
\newcommand{\Dlr}{\buildrel \leftrightarrow \over D\raise-1pt\hbox{}}

\usepackage{amsmath}
\usepackage{amsfonts,amssymb}
\title{Nucleon generalized form factors with twisted mass fermions}

\ShortTitle{Nucleon generalized form factors with twisted mass fermions}

\author{\speaker{Constantia  Alexandrou }\\
         Department of Physics, University of Cyprus, P.O. Box 20537, 1678 Nicosia, Cyprus, and\\  
 Computation-based Science and Technology Research  
    Center, Cyprus Institute, 20 Kavafi Str., Nicosia 2121, Cyprus \\  
        E-mail: \email{alexand@ucy.ac.cy}}

\author{Martha Constantinou\\
       Department of Physics, University of Cyprus, P.O. Box 20537, 1678 Nicosia, Cyprus\\
        E-mail: \email{constantinou.martha@ucy.ac.cy}}
\author{Vincent Drach\\
       NIC, DESY, Platanenallee 6, D-15738 Zeuthen, Germany\\
        E-mail: \email{vincent.drach@desy.de}}
\author{Karl Jansen\\
       NIC, DESY, Platanenallee 6, D-15738 Zeuthen, Germany\\
        E-mail: \email{karl.jansen@desy.de}}
\author{Christos Kallidonis\\
        Department of Physics, University of Cyprus, P.O. Box 20537, 1678 Nicosia, Cyprus, and\\  
 Computation-based Science and Technology Research  
    Center, Cyprus Institute, 20 Kavafi Str., Nicosia 2121, Cyprus\\
        E-mail: \email{kallidonis.christos@ucy.ac.cy}}
\author{Giannis Koutsou\\
 Computation-based Science and Technology Research  
 Center, Cyprus Institute, 20 Kavafi Str., Nicosia 2121, Cyprus\\
        E-mail: \email{g.koutsou@cyi.ac.cy}}

\abstract{
We present results on the nucleon form factors, momentum fraction and  helicity
moment for $N_f=2$ and $N_f=2+1+1$ twisted mass fermions for a number of lattice volumes and lattice spacings.  First results  for a new $N_f=2$ 
ensemble at the physical pion mass are also included.   The
  implications of these results on the spin content of the nucleon
are discussed  taking into account the disconnected contributions
at one pion mass.}

\FullConference{31st International Symposium on Lattice Field Theory - LATTICE 2013\\
		July 29 - August 3, 2013\\
		Mainz, Germany}

\begin{document}

\section{Introduction}
Fundamental properties of the nucleon such as its charge radius, magnetic moment
and axial charge have been studied experimentally for over 50 years.
This work focuses in computing these fundamental properties within the
lattice QCD formulation using twisted mass fermions (TMF). Several $N_f=2$ and $N_f=2+1+1$ ensembles generated at three different
lattice spacings smaller than 0.1~fm ~\cite{Boucaud:2007uk,Baron:2010bv} are analyzed~\cite{Alexandrou:2013joa}. The twisted mass formulation
is particularly suited for hadron structure calculations since it provides automatic ${\cal O}(a^2)$ improvement requiring no operator modification~\cite{Frezzotti:2003ni}. Our results include a simulation using the Iwasaki gluon action, and $N_f=2$ TMF with a clover term at the physical value of the pion mass, referred to as the physical ensemble~\cite{Kostrzewa:2013}. 

\section{Setting scale}
\begin{figure}[h!]
\begin{minipage}{0.49\linewidth}\vspace*{-0.3cm}
\includegraphics[width=\linewidth]{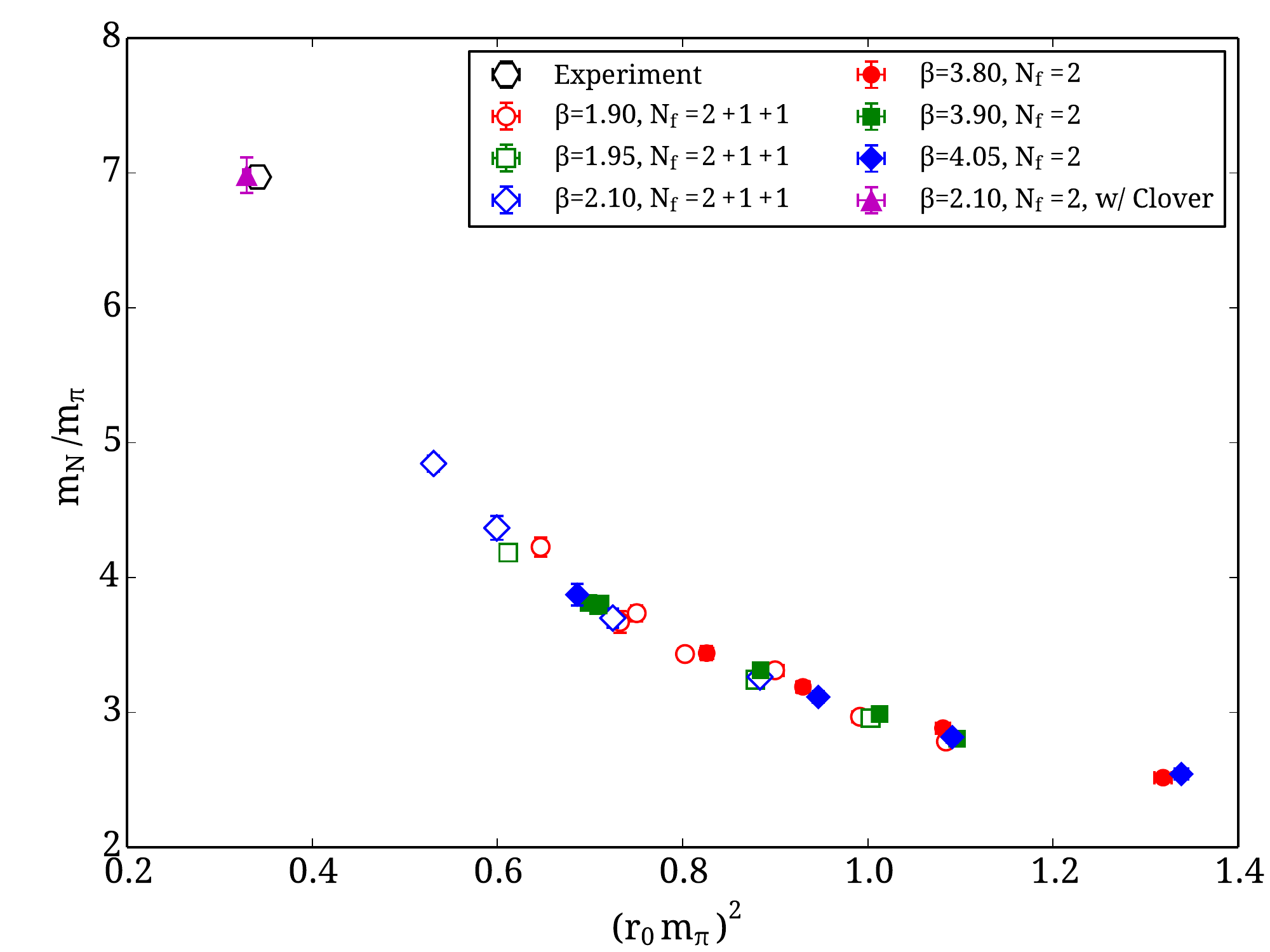}
\end{minipage}\hfill
\begin{minipage}{0.49\linewidth}
\includegraphics[width=\linewidth, height=0.8\linewidth]{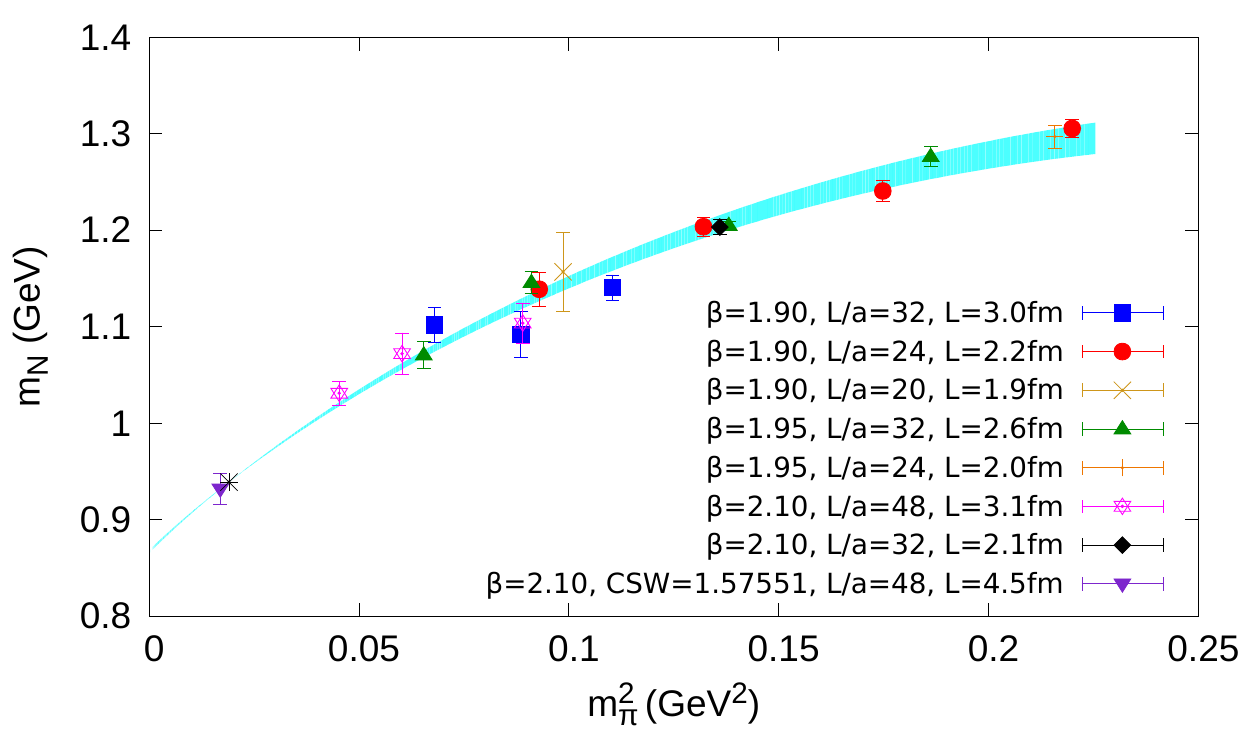}
\end{minipage}
\caption{Left: The ratio of the nucleon mass to the pion mass versus $(m_\pi r_0)^2$. Right: Chiral extrapolation of the nucleon mass to extract the lattice spacings.}\label{fig:mN}
\end{figure}

 For baryon observables we opt to set the scale by using the nucleon mass at the physical limit. In Fig.~\ref{fig:mN} we collect all the data on the nucleon mass, including our new result using $N_f=2$ the  physical ensemble. As it can be seen,
all data fall nicely on a universal curve.

The physical pion mass is expressed in units of $r_0$ the value of which
is  determined from the nucleon mass. Restricting to 
the $N_f=2+1+1$ ensembles generated with pion mass less than 300~MeV and 
including  the physical ensemble we find $r_0=0.495(6)$~fm.
In order to fix the lattice spacing $a$, we make a combined fit to the $N_f=2+1+1$
ensembles at $\beta=1.9$, $\beta=1.95$ and $\beta=2.1$, and to the physical ensemble using  the well-established  baryon chiral perturbation theory result
$ m_N = {m_N^0}- 4 {c_1}m_\pi^2 -\frac{3 g_A^2 }{16\pi f_\pi^2} m_\pi^3 $. 
The systematic error due to the chiral extrapolation is estimated  in two ways:
i) using the next order result in heavy baryon chiral perturbation theory (HB$\chi$PT) that includes explicit $\Delta$-degrees of freedom, and  ii)  varying the pion mass range for the fit. 
In performing a combined fit we assume that cut-off effects are negligible. By fitting the results at each $\beta$-value separately we obtain consistent values, thus verifying the smallness of cut-off effects.
 From the combined fit, we obtain
 $a=0.0936(13)(25)$~fm, $a=0.0823(11)(35)$~fm, $a=0.0646(7)(25)$~fm for $\beta=1.90$, 1.95 and 2.10 for the $N_f=2+1+1$ ensembles and $a=0.0937(2)(2)$~fm for our $N_f=2$ physical ensemble.  The lowest order chiral fit is shown in Fig.~\ref{fig:mN} and  describes well all lattice QCD data. The lattice spacings
for the $N_f=2$ ensembles were similarly determined and given in Ref.~\cite{Alexandrou:2010hf}.
The Feynman-Hellman theorem relates the coefficient $c_1$ of the $p^3$ fit to the $\sigma_{\pi N}$-term. Using the value of $c_1$ from our combined fit we find  $\sigma_{\pi N}=58(8)(7)$~MeV~\cite{Alexandrou:2011iu}. 

\vspace*{-0.3cm}

\section{High precision study of nucleon observables at $m_\pi=373$~MeV}
In order to study excited state contributions,  
we perform a  high-statistics study
using one  $N_f=2+1+1$ ensemble at $\beta=1.95$ and pion mass $m_\pi=373$~MeV, referred to as B55.32. 
We focus on three 
observables chosen because they  show a different degree of excited states contamination, namely the 
nucleon axial charge $g_A$ where they are expected to be small, the isovector momentum fraction $\langle x \rangle_{u-d}$ where we expect them
to contribute~\cite{Dinter:2011sg} and the
 scalar charge or equivalently the $\sigma$-terms where we expect  severe contamination~\cite{Alexandrou:2013nda}.
We also use this ensemble to compute the 
 disconnected contributions to all nucleon observables
with techniques reported in Ref.~\cite{Alexandrou:2013wca}.

As customary, we construct an appropriate ratio of the three-point function
we are interested in to 
two point functions, which for zero momentum transfer is given by $R(t_s,t_{\rm ins})=\frac{G^{\rm 3pt}(\Gamma^\mu, \> t_s,\>t_{\rm ins})}{G^{\rm 2pt}(\Gamma^0,\>t_s)}$.
We perform two types of  analysis for  the extraction of the matrix element: i) In the so-called plateau method we study  the large Euclidean time evolution of
the ratio 
 \be
    R(t_s,t_{\rm ins}) \xrightarrow[(t_s-t_{\rm ins})\Delta \gg 1]{(t_{\rm ins})\Delta \gg 1} \mathcal{M}[1
      + \dots e^{-\Delta({\bf p})t_{\rm ins}} + \dots e^{-\Delta({\bf p}^\prime)(t_s-t_{\rm ins})}]
\label{ratio}
\ee
where  $\mathcal{M}$ is the desired matrix element, $t_s$ and $t_{\rm ins}$, the
  sink and insertion separation time (we have taken the time of the source $t_0=0$), 
and $\Delta({\bf p})$ the
  energy gap between the first excited state and the ground state.
In the second approach  the ratio is summed  over $t_{\rm ins}$:
 \be
  \sum_{t_{\rm ins}=t_0}^{t_s} R(t_s,t_{\rm ins}) = {\sf Const.} + \mathcal{M}\>[t_s + \mathcal{O}(e^{-\Delta({\bf p})\>t_s})  + \mathcal{O}(e^{-\Delta({\bf p'})\>t_s})].
\label{summation}
\ee
In this so-called summation method~\cite{Maiani:1987by}, excited state contributions are suppressed by exponentials decaying with $t_s$, rather than $t_s-t_{\rm ins}$ and $t_{\rm ins}$.  However, one needs to fit the slope of the summed ratio rather than to a constant as in the plateau method.
We note that this result also holds if one does not include $t_0$ and $t_s$ in the sum, avoiding  contact term contributions. All results shown here
  do not include these terms.
 We use the incremental eigCG algorithm~\cite{Stathopoulos:2007zi} to speed-up the inversions at different $t_s$ yielding a factor of 3 speed-up. We note that with
one sequential inversion for each $t_{s}$ we obtain results 
for all operator insertions.

{\bf The axial charge $\mathbf{g_A}$:}
The nucleon axial charge $g_A$, defined as
the nucleon matrix element of the axial-vector current
$A^3_\mu=\bar{\psi}\gamma_\mu\gamma_5 \frac{\tau^3}{2}\psi$ at zero momentum transfer, is well-known experimentally and, being an isovector, receives  no quark loop contributions. Therefore, it can be considered as the simplest baryon observable beyond the mass.
\begin{figure}[h!]
 \begin{minipage}[t]{0.45\linewidth}
{    \includegraphics[width=\linewidth]{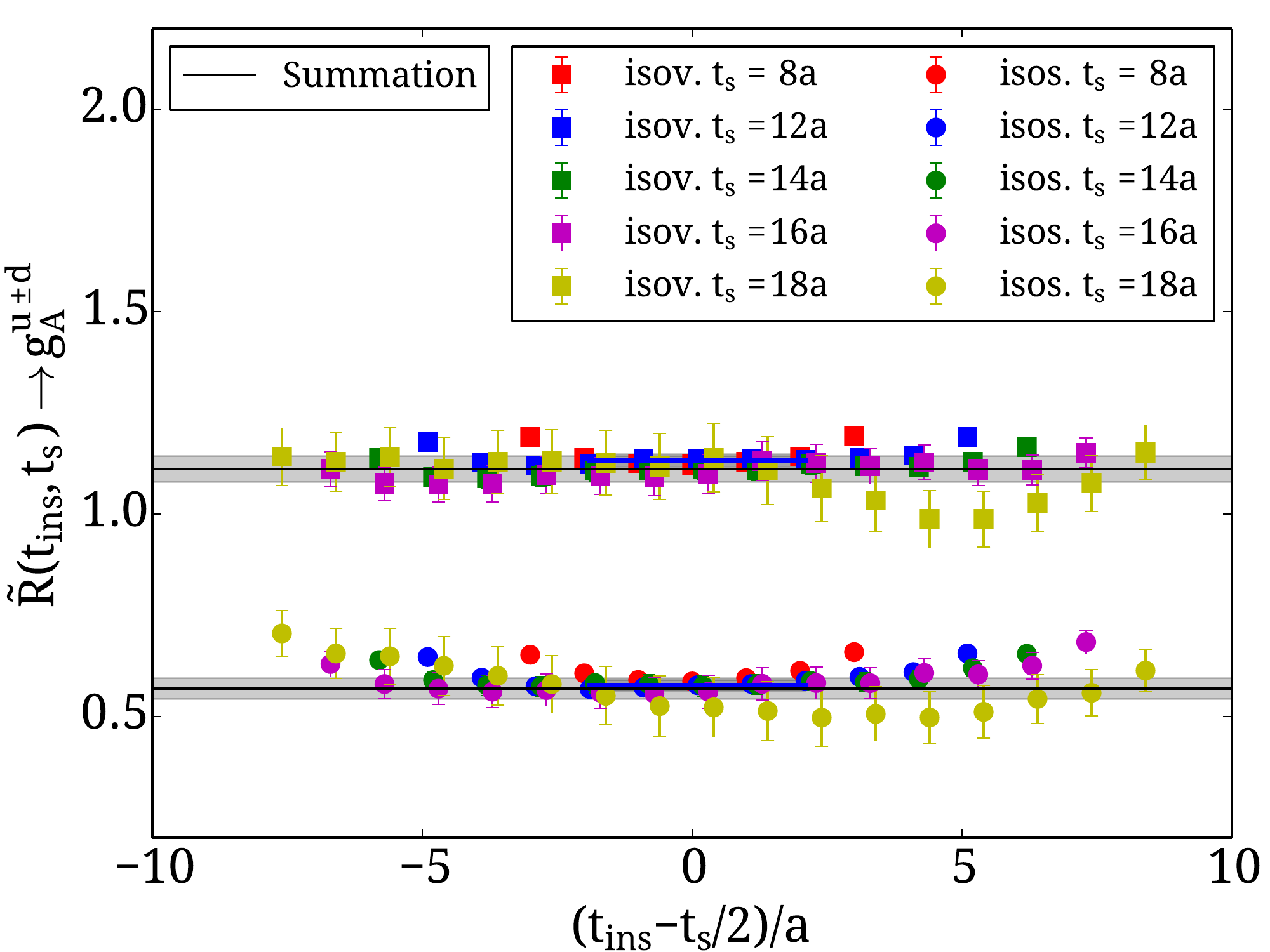}}
  \end{minipage}
  \hfill
  \begin{minipage}[t]{0.45\linewidth}
{\includegraphics[width=\linewidth]{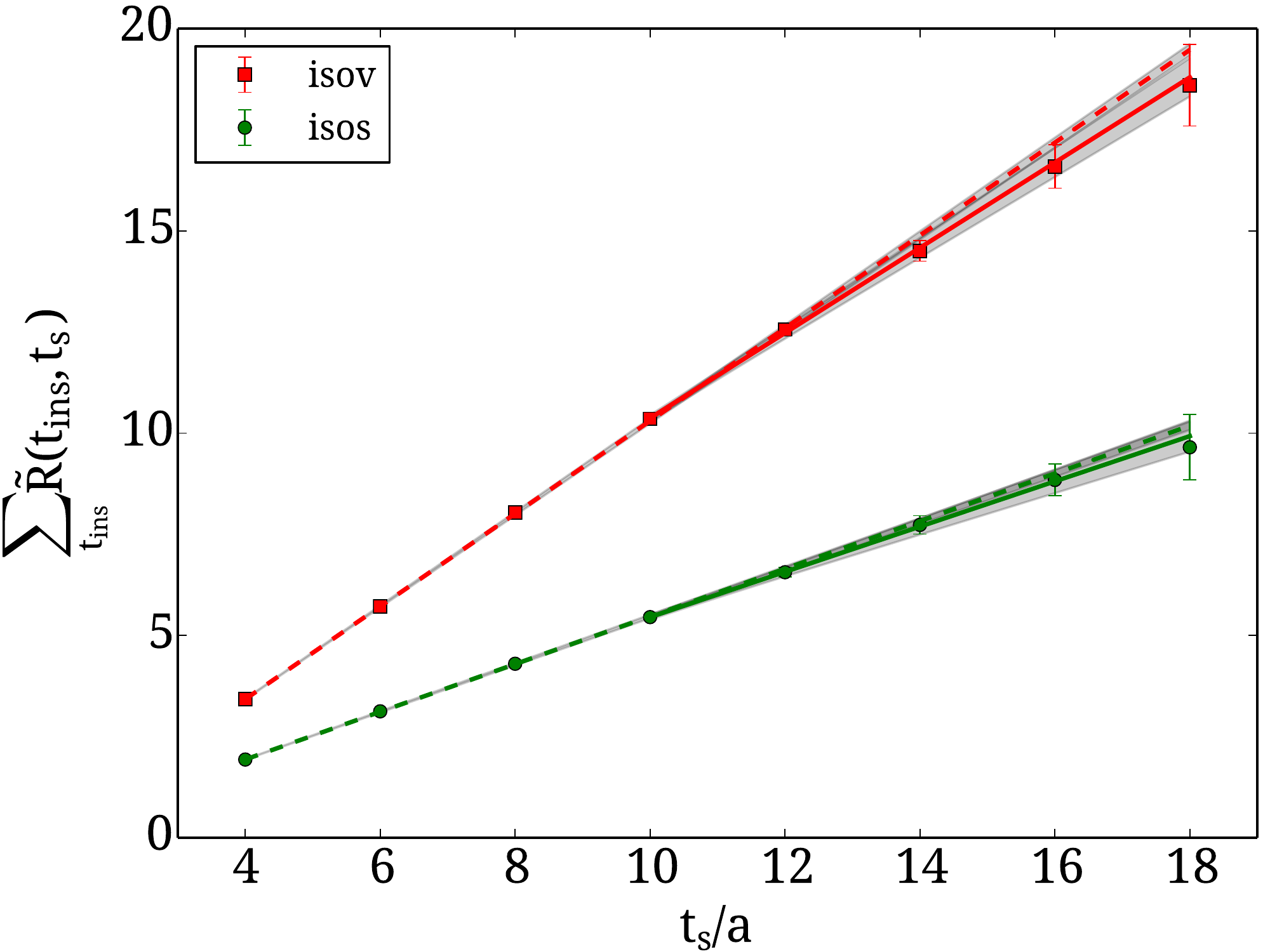}}
  \end{minipage}
\caption{Left: $\tilde{R}(t_{\rm ins},t_s)$ for the isovector (squares) and connected isoscalar (circles) axial charge. The blue bands show the plateau-value  for $t_s/a=12$. The gray bands are the results extracted by fitting to the slope of the summed ratio as shown in the right panel. A total of 1200 measurements were used for each $t_s$.}\label{fig:gA test}
\end{figure}
In Fig.~\ref{fig:gA test} we show the renormalized ratios $\tilde{R}(t_{\rm ins},t_s)$ from which  the isovector, and  the isoscalar axial charge
are determined for
the B55.32 ensemble. We observe that
  the value of the plateau is the same  for several $t_s$ or equivalently no curvature 
is seen in the summed ratio, which yields the same slope  either using initial fit time $t_i/a=4$ or $t_i/a=10$. This means that there is no detectable excited states contamination in the ratio yielding  $g_A$ even for $t_i/a=4$,
a result that corroborates our previous high precision study using the fixed current sequential method~\cite{Dinter:2011sg}.

\begin{figure}[h!]
  \begin{minipage}[t]{0.45\linewidth}
    \includegraphics[width=\linewidth]{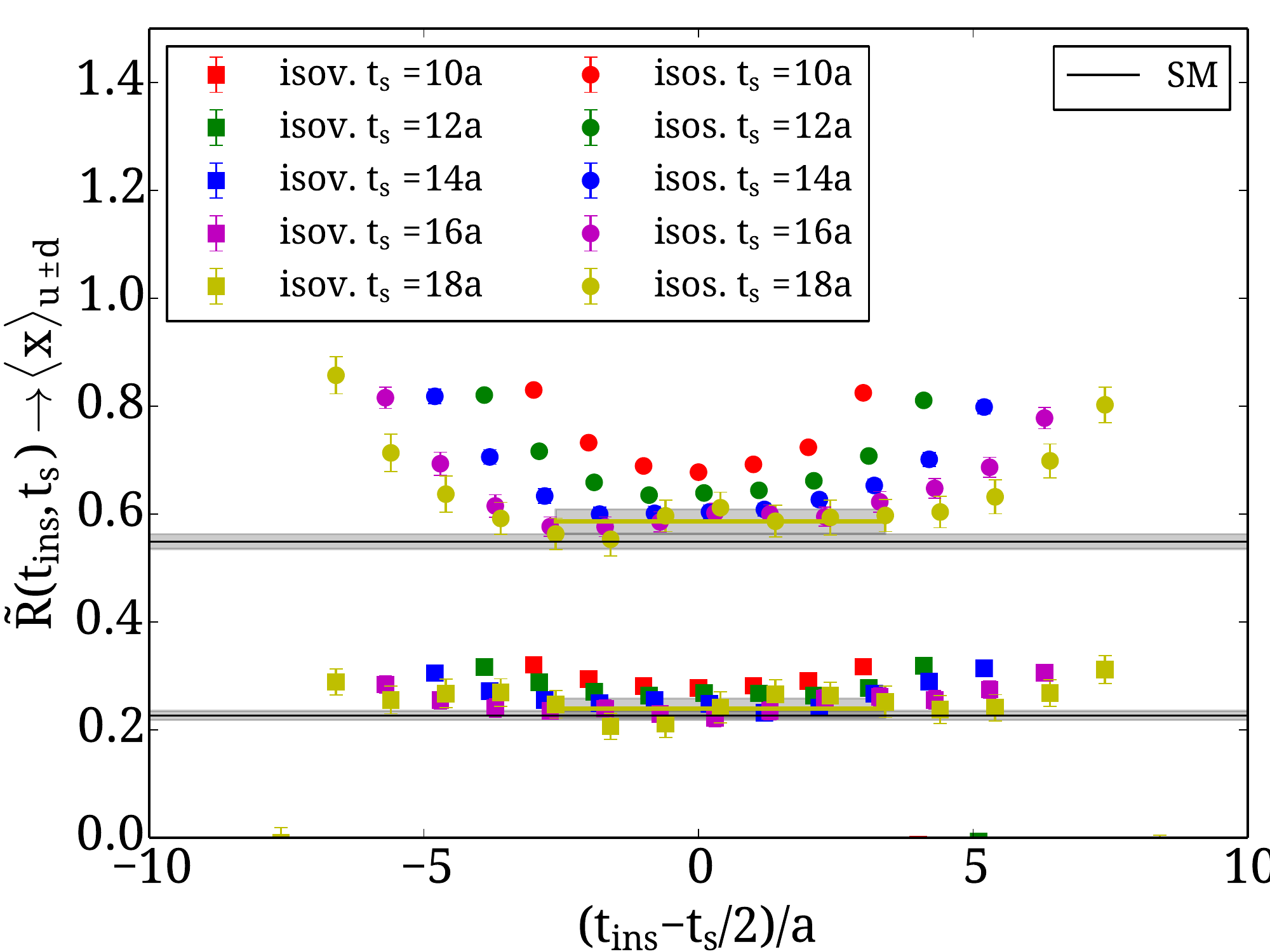}
  \end{minipage}
  \hfill
  \begin{minipage}[t]{0.45\linewidth}
 { \includegraphics[width=\linewidth]{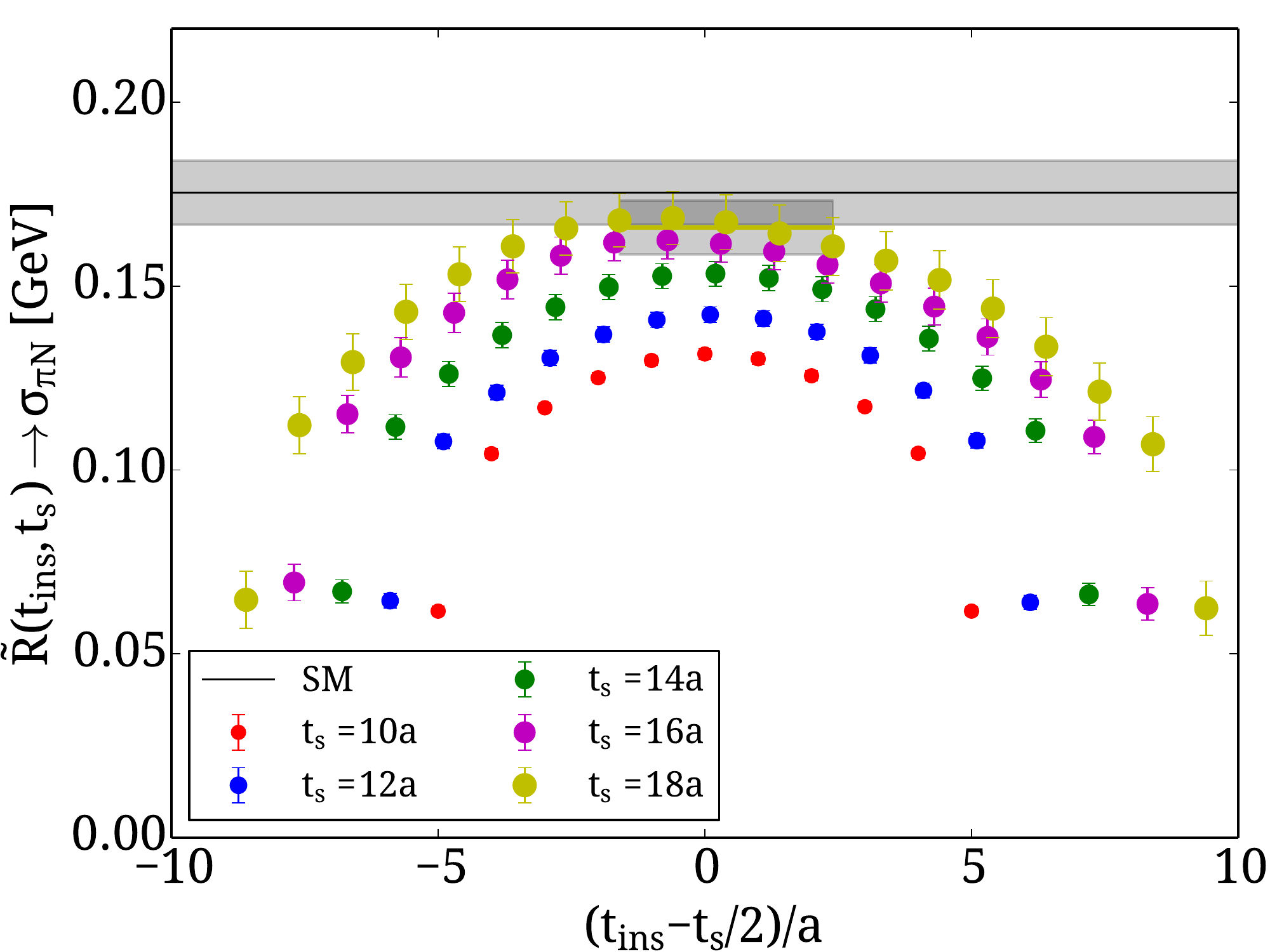}}
  \end{minipage}
\caption{The ratio from which the isovector and connected isoscalar momentum fraction is extracted (left) and the corresponding ratio for $\sigma_{\pi N}$ (right). The notation is the same as that of Fig.~2. 
}
\label{fig:x test}
\end{figure}

{\bf The quark momentum fraction and the scalar charge:}
An analogous study is carried out for the quark  momentum fraction $\langle x \rangle_{u-d}$ extracted from the nucleon matrix elements of $ {\cal O}^{\mu_1 \mu_2}  = \bar \psi  \gamma^{\{\mu_1}i\Dlr  ^{\mu_{2}\}} \psi$ at 
momentum transfer squared $q^2=0$. 
Our results for the B55.32 ensemble are shown in Fig.~\ref{fig:x test}.  The renormalized ratio for $\langle x \rangle_{u-d}$ decreases as the sink-source separation increases. Fitting the plateau when $t_s/a=18$ yields consistent results  with those obtained from the summation method, for both
the isovector and isoscalar quantities. Such agreement is what is  expected when excited
state contributions become negligible.
The isoscalar charge  $g_s^{u+d} = \langle N|{\bar u}u+\bar{d}d|N\rangle$ is computed in an  analogous way to the calculation of the nucleon sigma-terms making use of the advantages of twisted mass fermions~\cite{Alexandrou:2013nda,Dinter:2012tt}. The scalar and tensor charges,  $g_s$ and $g_T$, provide constrains for possible scalar and tensor interactions at the TeV scale~\cite{Bhattacharya:2013ehc}.
 In Fig.~\ref{fig:x test} we show the
ratio from which the $\sigma_{\pi N}$ or equivalently the isoscalar scalar charge   is extracted.
Fitting at $t_s\sim 1.0$~fm will underestimate the scalar charge or $\sigma_{\pi N}$-term by 25\%. Fitting the plateau and the summed ratio at large enough time separations  yields consistent results.  Thus consistency between the plateau and the summation method is a prerequisite for ascertaining  that  excited states contributions to
the ratio are negligible.

\begin{figure}[h!]
 \  \begin{minipage}[t]{0.33\linewidth}
  \includegraphics[width=\linewidth]{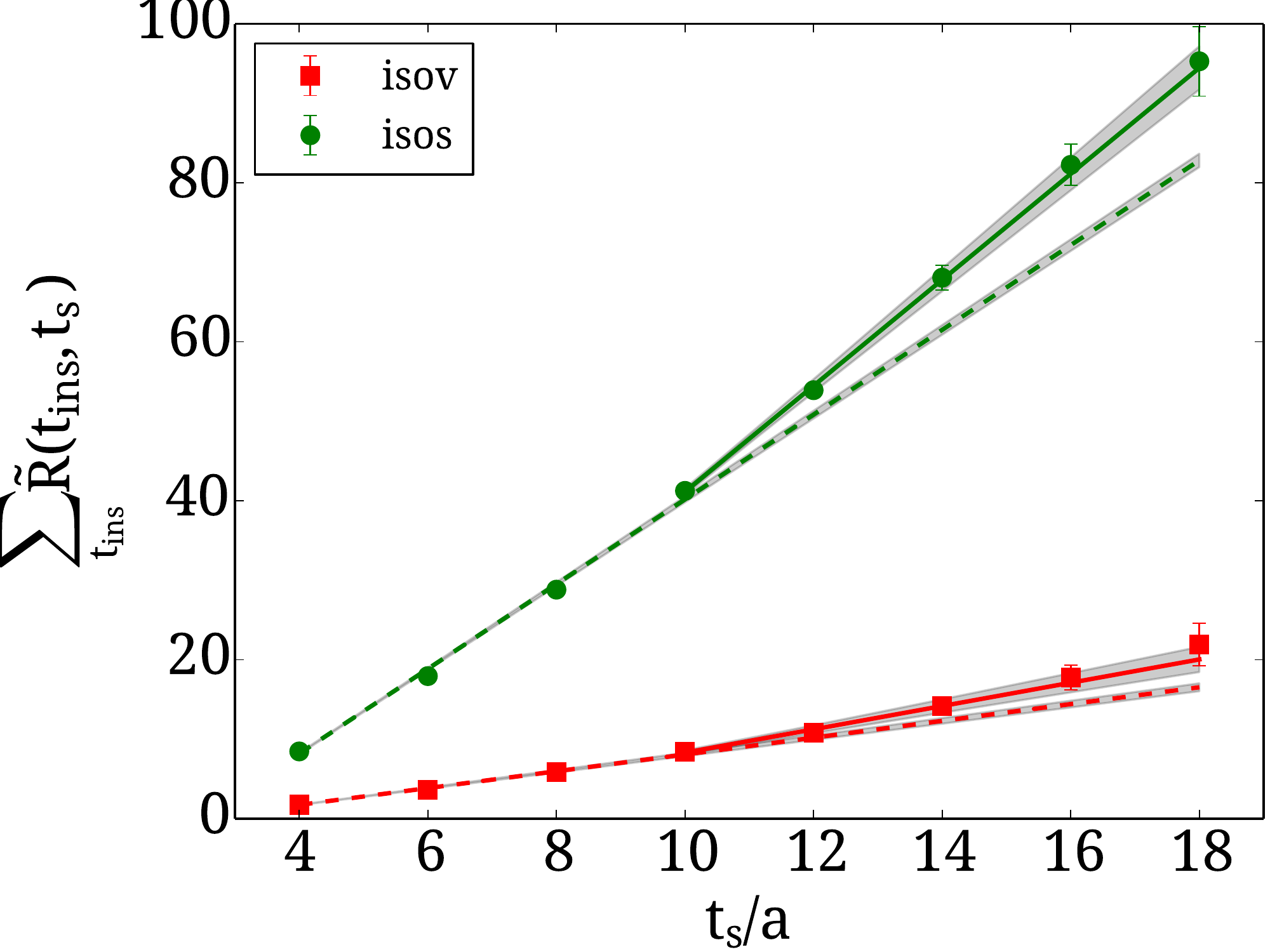}
 \end{minipage}
  \hfill
  \begin{minipage}[t]{0.33\linewidth}
   \includegraphics[width=\linewidth]{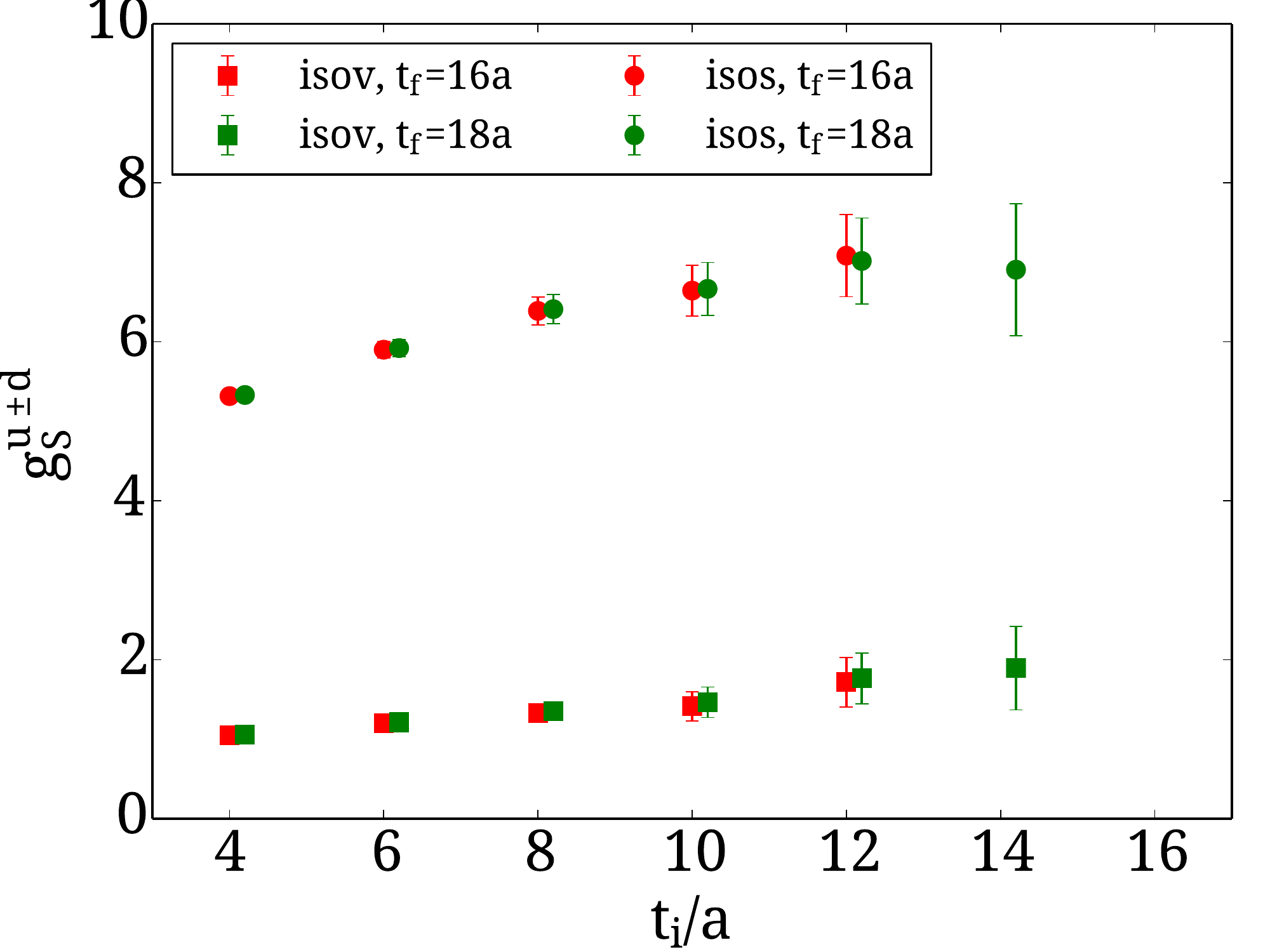}
  \end{minipage}\hfill
\begin{minipage}[t]{0.32\linewidth}
  \includegraphics[width=\linewidth]{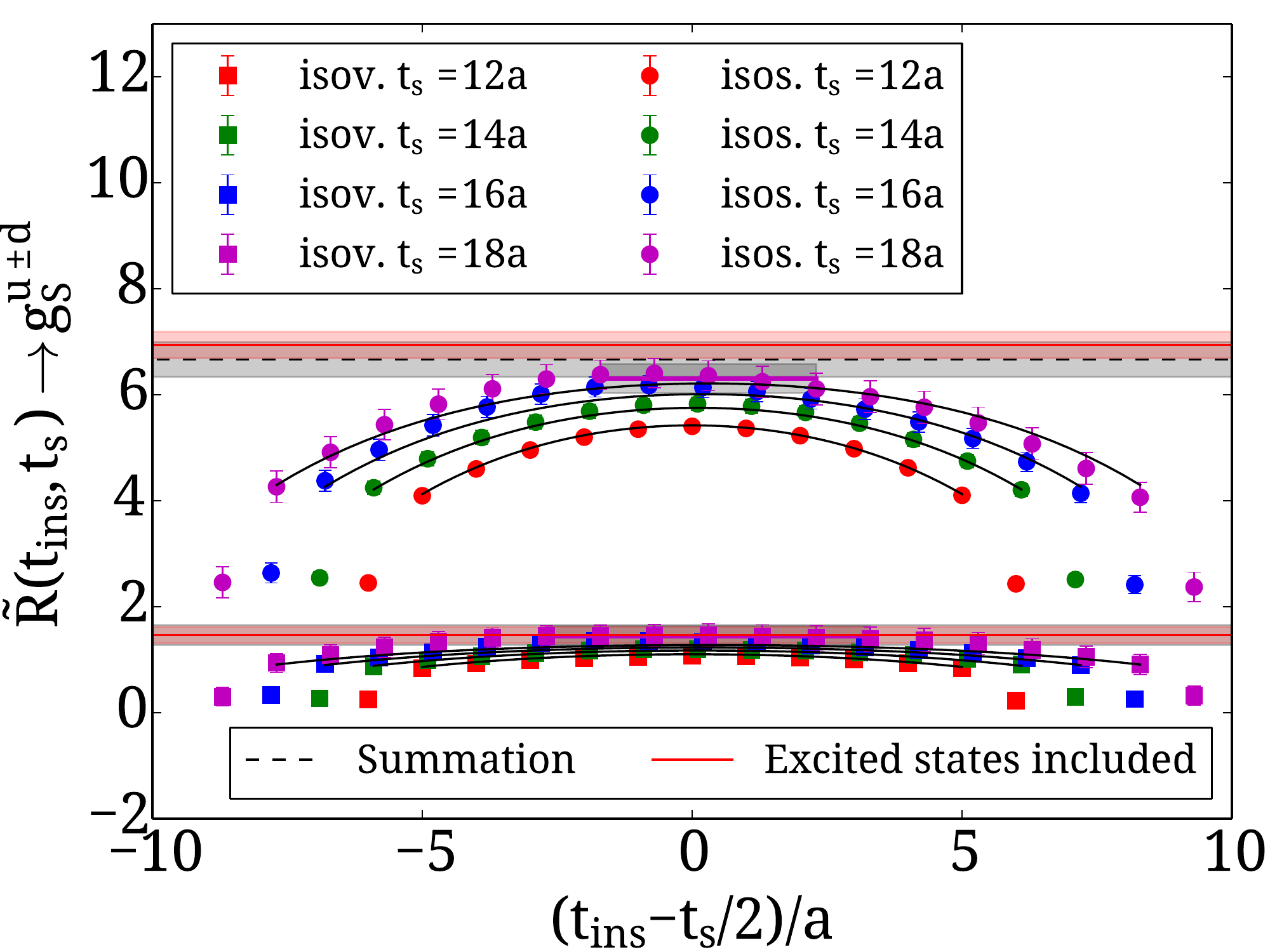}
  \end{minipage}
\caption{The summed renormalized ratio (left) and the slope (center) as a function of the initial fitting time $t_i/a$
for two final fitting ranges $t_f/a=16$ and $t_f/a=18$. The solid lines
in the right panel show the combined fit to results obtained for a number of $t_s$  that explicitly includes the contributions of the first excited state. }\label{fig:gs test}
\end{figure}
In Fig.~\ref{fig:gs test} we examine further the excited state contributions to the scalar charge. We show the dependence of the slope extracted from the summed renormalized ratio on the lower fit range $t_i/a$.  Different
values are obtained if we perform the fit using $t_i/a=4$ as compare to using $t_i/a=8$.
Thus,  just like in the plateau method, one seeks convergence of the slope as $t_i$ increases, as shown in the center-plot in Fig.\ref{fig:gs test}. 
The value of the slope is not affected when changing the upper fitting range $t_f/a$ from 16 to 18.
 We find that we need  $t_i/a \ge 10$ to damp sufficiently  excited state contributions. One can also  include explicitly in the fitting function 
 the terms due to the excited states. Taking into account the contributions of the first excited state
and making a combined fit to all time separations using four fitting parameters
yields consistent results, as demonstrated in Fig.~\ref{fig:gs test}. We 
stress that all methods to probe excited states  require an evaluation of the three-point function for several sink-source time separations.
For the scalar charge and $\sigma_{\pi N}$, one observes
severe contamination from excited states requiring sink source separations of at least 1.5~fm. Agreement of summation, plateau and two-state fits give confidence to the correctness of the final result.

The isoscalar axial, momentum fraction  and scalar charge have disconnected contributions, which
need special techniques for their computation. We have calculated these contributions in Ref.~\cite{Alexandrou:2013wca} for the B55.32 ensemble using  $\sim$150,0000~statistics on 4700 confs. We have found a non-zero result for $g_A$ and $g_s^{u+d}$, which is about 10\% of the connected part, whereas for $\langle x\rangle_{u+d}$ the result is conisistent with zero giving an upper bound on the size of the
disconnected contribution. 

\vspace*{-0.3cm}

\section{Results for $130$~MeV $<m_\pi <450$~MeV}

\vspace*{-0.3cm}

In this section we show results using a number of TMF ensembles for the axial charge,
momentum fraction and scalar charge, including results for the physical ensemble.

\begin{figure}[h!]\vspace*{-0.3cm}
\begin{minipage}{0.49\linewidth}
\includegraphics[width=\linewidth]{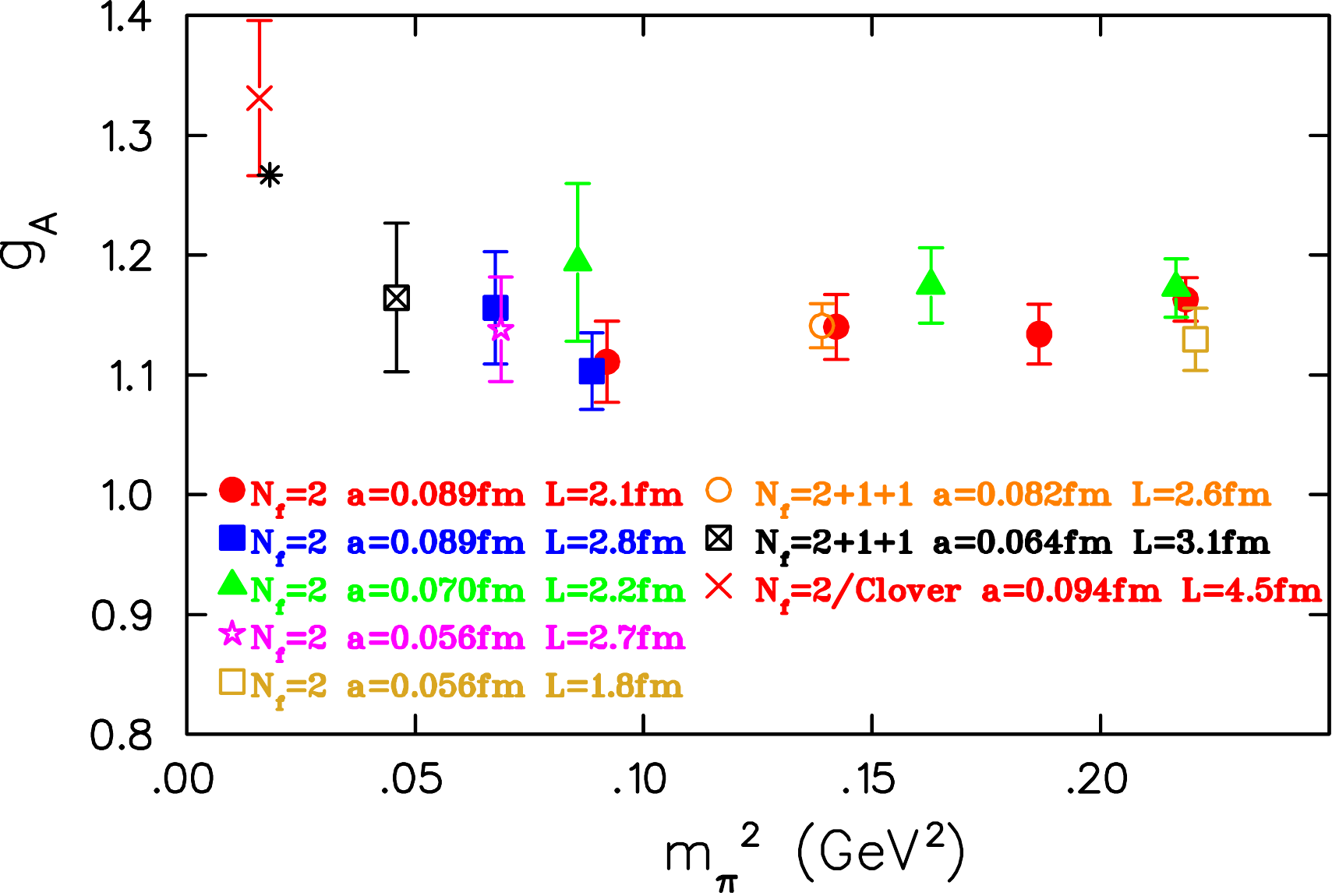}
\end{minipage}\hfill
\begin{minipage}{0.49\linewidth}
{\includegraphics[width=\linewidth]{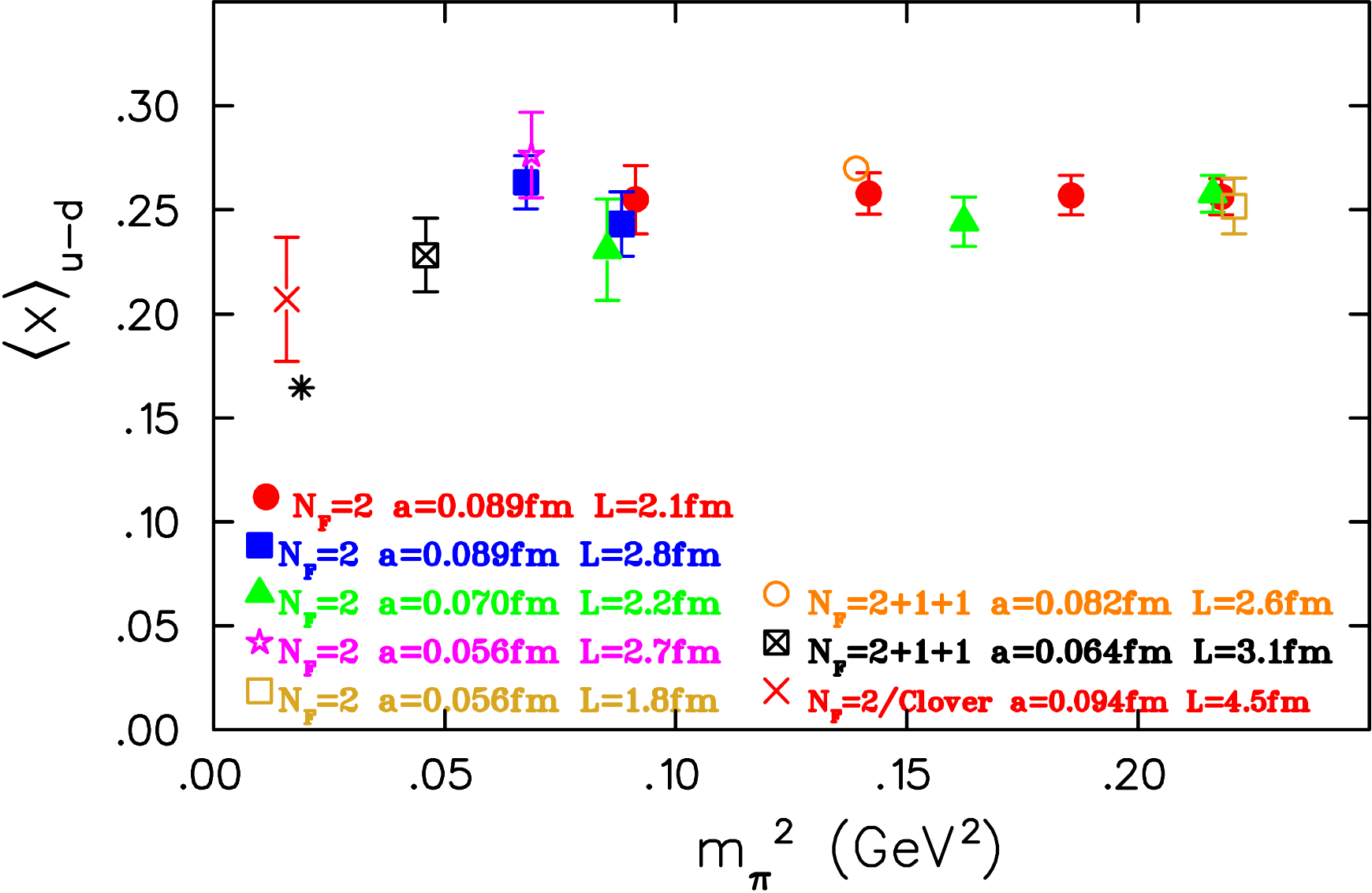}}
\end{minipage}
\vspace*{-0.3cm}
\caption{TMF results on several ensembles for the axial charge (left) and $\langle x\rangle_{u-d}$~\cite{Alexandrou:2013joa} (right).}\label{fig:gA}
\end{figure}

As can be seen in Fig.~\ref{fig:gA}, where  results
for $g_A$ and  $\langle x\rangle_{u-d}$ are displayed,  our values at physical pion mass are consistent with the 
experimental ones, albeit with still large statistical error.   A number of collaborations  are currently engaging in systematic studies using simulations at near physical pion masses~\cite{Green:2012ud,Horsley:2013ayv,Capitani:2012gj,Owen:2012ts,Bhattacharya:2013ehc} and we therefore  expect that the values of these
key observables at the physical point will soon be more accurately determined.
Having lattice results on the axial charge and the first moment of the unpolarized
distribution  we can examine the spin carried by quarks $J^q=\frac{1}{2}(A_{20}^q(0)+B_{20}^q(0))$.
Both  $A_{20}^q(Q^2)$ and $B_{20}^q(Q^2)$  are extracted from the nucleon matrix elements of $ {\cal O}^{\mu_1 \mu_2}  = \bar \psi  \gamma^{\{\mu_1}i\Dlr  ^{\mu_{2}\}} \psi$ in the $\overline{MS}$ scheme at  $\mu= 2$~GeV using non-perturbative renormalization.

\begin{figure}[h!]
\begin{minipage}{0.33\linewidth}
{\includegraphics[width=\linewidth]{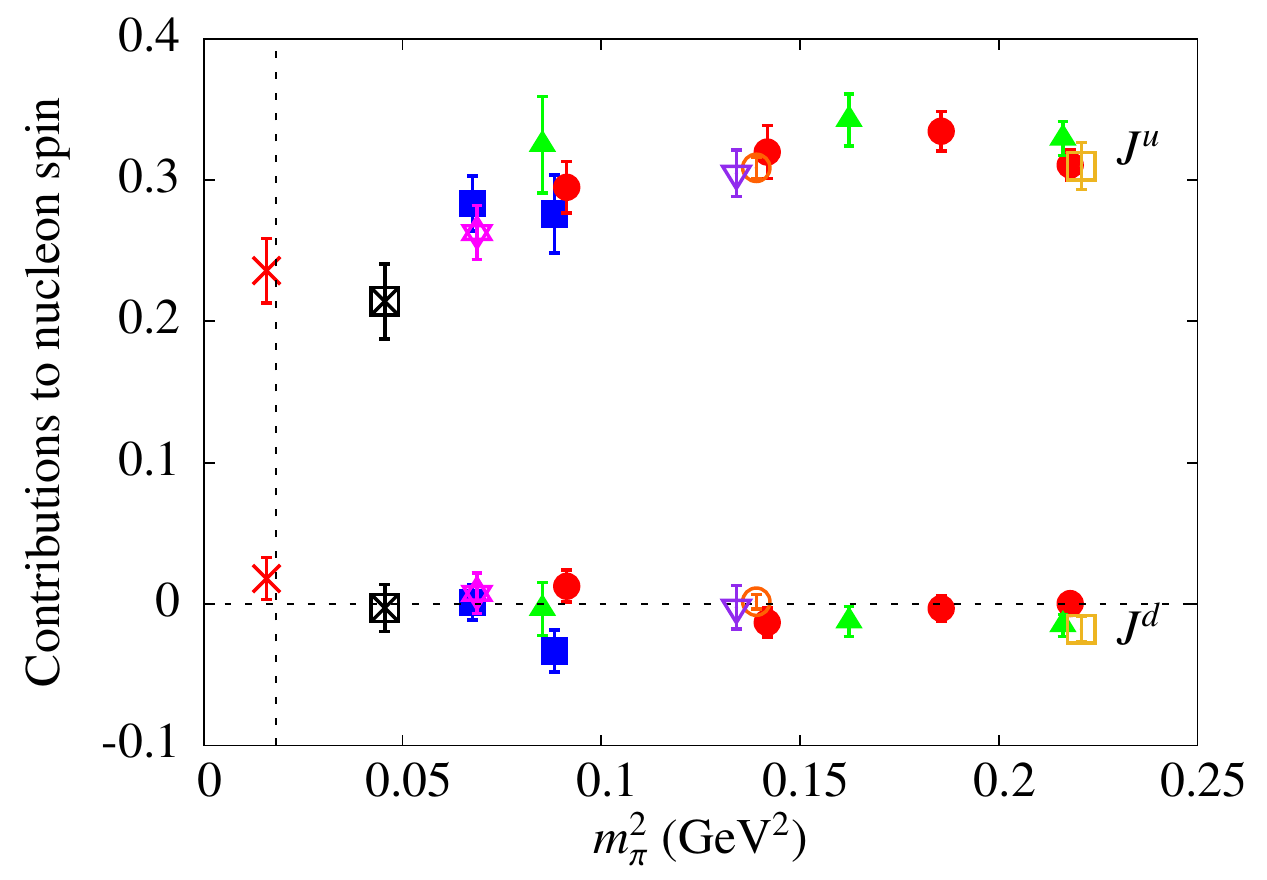}}
\end{minipage}\hfill
\begin{minipage}{0.33\linewidth}
{\includegraphics[width=\linewidth]{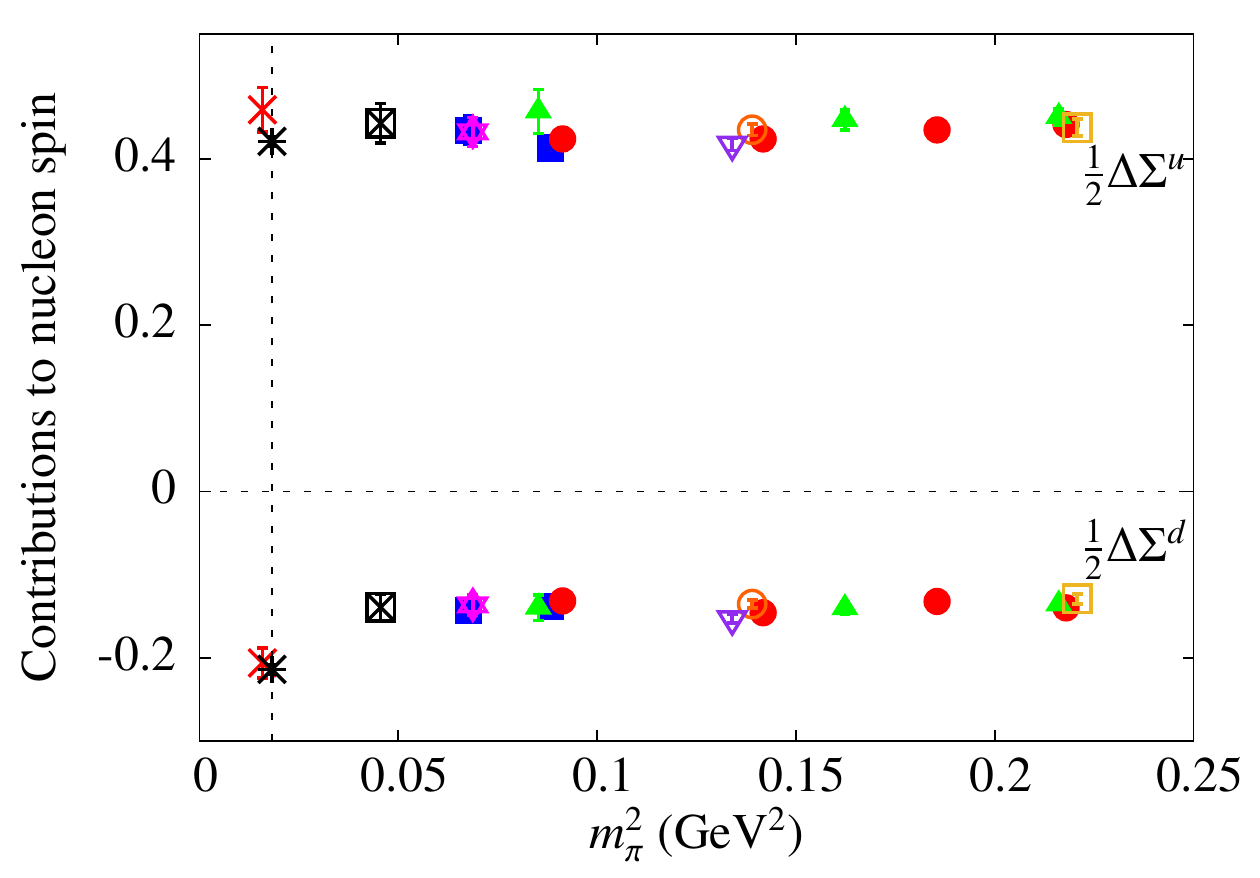}}
\end{minipage}
\begin{minipage}{0.33\linewidth}
{\includegraphics[width=\linewidth]{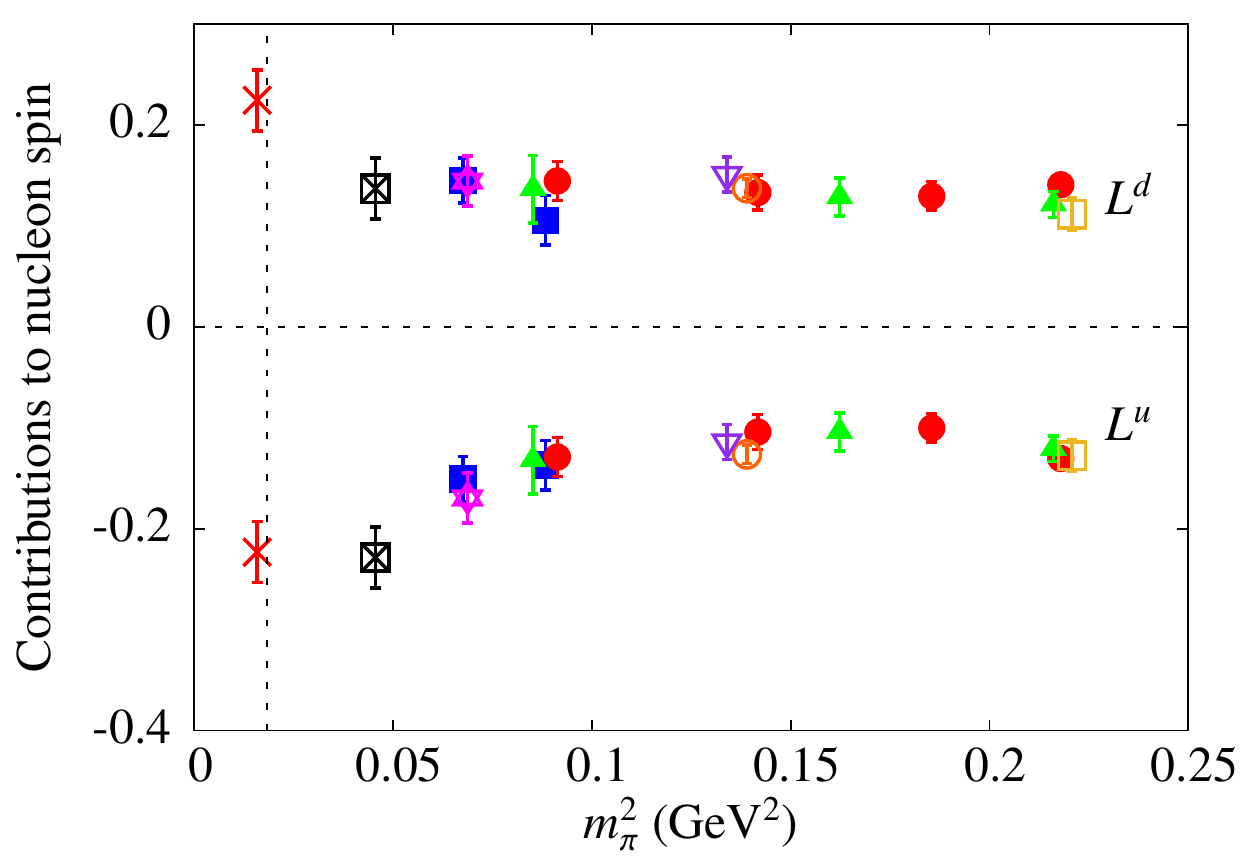}}
\end{minipage}
\label{fig:spin}
\caption{The spin $J^q=\frac{1}{2}\Delta \Sigma^q+L^q$ carried by the u and d quarks neglecting disconnected contributions except for  B55.32 at $m_\pi=373$~MeV, which includes the disconnected part (purple downward triangle). }
\end{figure}

 \begin{figure}[h!]
\begin{minipage}{0.49\linewidth}
  \includegraphics[width=\linewidth]{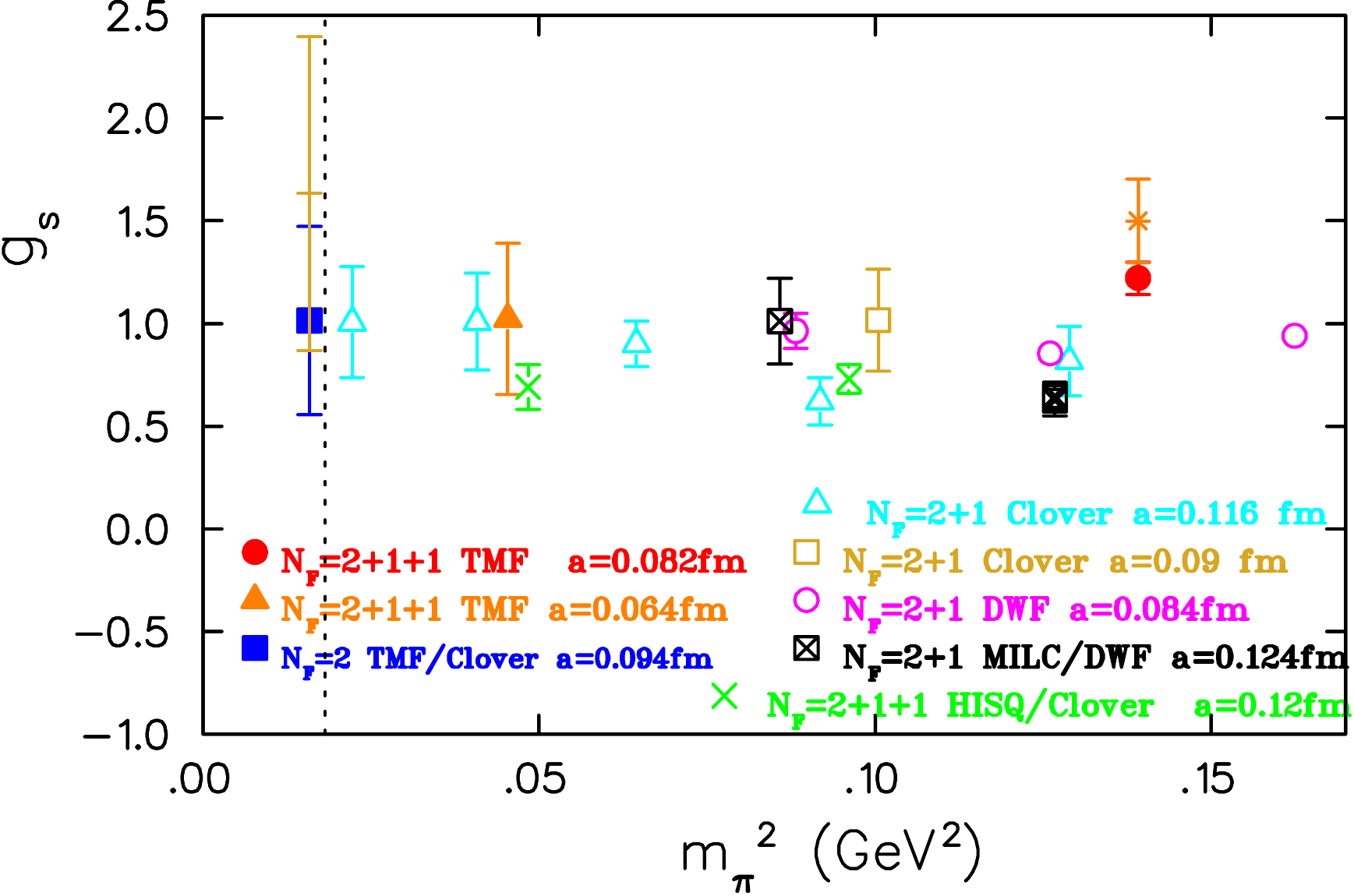}
  \end{minipage}
\begin{minipage}{0.49\linewidth}
\includegraphics[width=\linewidth]{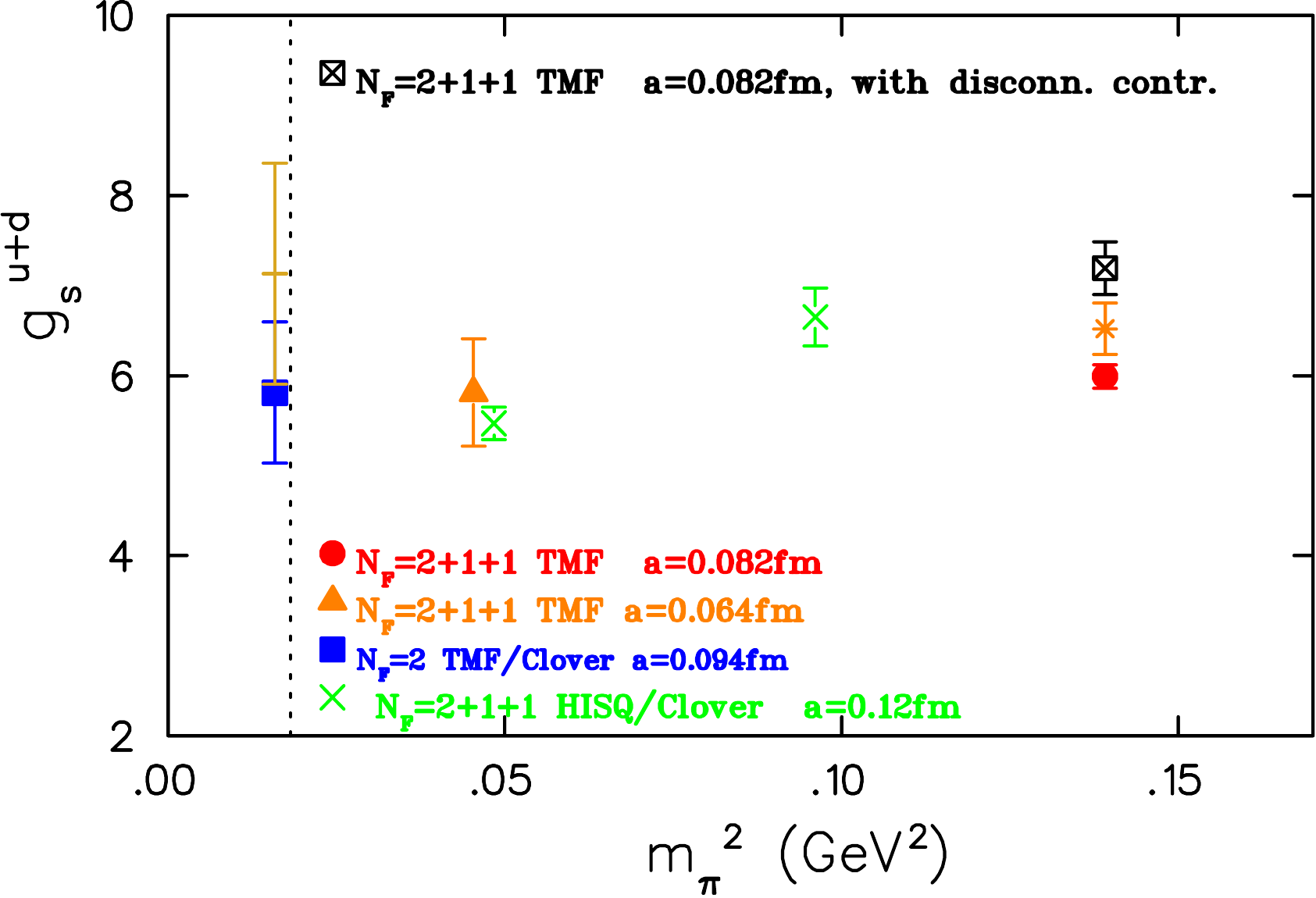}
\end{minipage}
\vspace*{-0.3cm}
\caption{The isovector (left) and isoscalar (right) scalar charge. The orange star shows the result when the sink-source time separation is increased from 1.2~fm (filled circle) to $\sim 1.5$~fm for the B55.32 ensemble.  The two points at the physical pion mass are obtained for $t_s=1.1$~fm and 1.3~fm. The black square in the plot on the right panel  
includes the disconnected contribution.}\label{fig:scalar}
\end{figure}

In Fig.~\ref{fig:spin}  we show the spin and
angular momentum carried by the quarks in the nucleon. As can be seen, our
preliminary results at the physical point are in agreement with experiment.
A high statistics analysis of the disconnected contributions for the B55.32 ensemble shows a $\sim$10\% contribution on $\Delta \Sigma^{u+d}$, which decreases its value as
shown in Fig.~\ref{fig:spin}, an effect that tends to bring its value towards the
experimental one. In Fig.~\ref{fig:scalar} we show the scalar charge computed for a  sink-source time separation of 1~fm-1.2~fm.
However,
this observable has large excited state contamination and increasing $t_s/a$ to 1.5~fm increases 
its value as shown in Fig.~\ref{fig:scalar} for the B55.32 ensemble.
 The disconnected
contribution, also computed for the B55.32 ensemble, increases its value further.
It is thus important, in order to obtain a reliable result,  to perform the calculation at larger $t_s/a$ and include the disconnected part.

\vspace*{-0.3cm}

\section{Conclusions}

\vspace*{-0.3cm}

Simulations at the physical point are now becoming available enabling
  results on  $g_A$, $\langle x \rangle _{u-d}$ and other
interesting observables directly at the physical point. High statistics and careful cross-checks of lattice artifacts will be needed to finalize these  results  Noise reduction techniques such as the truncated solver method and all-mode-averaging will be crucial for future computations.
Evaluation of disconnected  quark loop diagrams has become feasible
and disconnected contributions must be taken into account  for observables such as the
axial and scalar charges.
Confirmation of experimentally known quantities such as $g_A$ will enable reliable predictions of other less well-measured observables  providing insight into the structure of hadrons and input that is crucial for  new physics, such as the value of the nucleon $\sigma$-terms, the scalar and tensor charges. 

\vspace*{0.3cm}

\noindent
{\bf Acknowledgements:}
Partial support was provided by the projects EPYAN/0506/08,
TECHNOLOGY/ $\Theta$E$\Pi$I$\Sigma$/0311(BE)/16 and 
$\Pi$PO$\Sigma$E$\Lambda$KY$\Sigma$H/EM$\Pi$EIPO$\Sigma$/0311/16 funded by the Cyprus Research Promotion Foundation, and by the EU ITN project
 PITN-GA-2009-238353 (ITN
STRONGnet). This work used computational resources provided by PRACE, JSC, Germany and the  Cy-Tera project (NEA Y$\Pi$O$\Delta$OMH/$\Sigma$TPATH/0308/31).

\end{document}